# The Melting Line of Hydrogen at High Pressures

Shanti Deemyad and Isaac F. Silvera

Lyman Laboratory of Physics, Harvard University, Cambridge MA, 02138

The insulator to metal transition in solid hydrogen was predicted over 70 years ago but the demonstration of this transition remains a scientific challenge. In this regard, a peak in the temperature vs. pressure melting line of hydrogen may be a possible precursor for metallization. However, previous measurements of the fusion curve of hydrogen have been limited in pressure by diffusion of hydrogen into the gasket or diamonds. To overcome this limitation we have used an innovative technique of pulsed laser heating of the sample and final peak in the melting line at P=$64.7 \pm 4$ GPa and T=$1055 \pm 20$ K.

One of the great challenges to condensed matter physics is the observation of the insulator to metal (IM) transition in solid hydrogen, predicted over 70 years ago by Wigner and Huntington to take place at 25 GPa [1]; metallic hydrogen was also predicted by Ashcroft [2] to be a high temperature superconductor. Earlier advances at low temperature found two new high pressure phases, the broken symmetry phase (BSP) [3, 4] at about 110 GPa and the hydrogen A-phase [5, 6] at about 150 GPa, but neither in the metallic phase [7, 8] . Pressures as high as 342 GPa [9, 10] have not yielded the metallic state. Recently the approach to finding the IM transition in diamond anvil cells (DACs) has shifted from low to very high temperature. After a calculation by Scandolo [11], an analysis of the melting line by Bonev et al [12] predicted a peak at ~80 GPa and ~900 K. Furthermore, at higher temperatures they found a liquid-liquid transition from $H_2$ to non-molecular hydrogen, with a negative slope. By extrapolation beyond the pressures of the peak this curve meets the melting line so that hydrogen would melt from the molecular solid to the atomic liquid. Extrapolation of this melt line to still higher pressure and lower temperature implies that at a pressure greater than 400 GPa hydrogen might be a liquid at T=0 K. Babaev, Sudbo, and Ashcroft [13] analyzed high pressure liquid atomic hydrogen in a magnetic field and found two-component superconductivity for the protons and electrons, as well as superfluidity. The existence of a peak in the melting line will show a possible new pathway to metallic hydrogen and the importance of extending calculations and experiments to higher pressures and lower temperatures. Using an innovative technique we have measured the melting line in the region where molecular hydrogen melts to a molecular liquid and find a peak at a pressure of $64.7 \pm 4$ GPa and temperature of $1055 \pm 20$ K.

In order to measure the melting line a hydrogen sample must be confined at high pressure and temperature and this is a challenge. Diatschenko and Chu [14] measured the pressure-temperature melting curve up to room temperature and 5.2 GPa and fit it to the conventional Simon-Glatzel equation [15]. Datchi, Loubeyre, and LeToullec [16] extended the curve in an electrically heated DAC to P=15 GPa and T= 526 K. This was their limiting temperature, since at higher temperatures the hydrogen diffused through the confining metallic gasket and the sample was lost. They found a curvature in the melting line such that it could be fit to a Kechin melting curve [17] which can have a maximum. Subsequently, Gregoryanz et al [18] were able extend the melting line measurements to P~44 GPa and T~800 K, also fitting to a Kechin melting curve, with a possible maximum at higher pressures. In their ohmically heated cell, at the highest temperatures the diamonds would fail within a few minutes, evidently due to hydrogen diffusion into the diamonds, leading to diamond embrittlement [19]. Thus, there is a need to extend the measurements to higher temperatures [20].

There are two common methods of heating samples in DACs: continuous (CW) electrical or ohmic heating of the DAC where the temperature is measured with a thermocouple, and CW laser heating. In the latter an optical absorber is embedded in the sample in the DAC. A high power laser (~50 Watts) is focused on the absorber; the absorber and the surrounding sample heat to thousands of degrees K, controlled by the laser power. Temperature can be accurately determined by recording the blackbody (BB) radiation spectrum and fitting to a Planckian BB curve. This method has been used, for example, to measure the melting lines of rare gas solids [21], where melting was detected by illuminating the sample with an auxiliary laser and observing the onset of motion in the laser speckle pattern. For ohmic heating the sample temperature is uniform; for laser heating there is a large thermal gradient. The absorbed power flows from the absorber to the sample to the diamonds to the body of the DAC, and components become hot and susceptible to hydrogen diffusion and embrittlement, while the DAC body can become hot to the touch. To circumvent the problems due to diffusion, we have utilized the method of pulsed laser heating [22-24]. We use a neodymium-vanadate laser with a pulse length of 70 to ~200 ns and peak powers up to ~20 KWatts. During the pulse the surface of the absorber, a 1.5 micron thick platinum foil with linear dimension ~25 microns (see Fig. 1) heats and the hydrogen pressed against its surface also heats to the same temperature [25]. The absorber was separated from the surface of the diamond anvil by grains of ruby, as shown in Fig. 1, so that it would not directly contact and heat the diamond surface.

Since the time required to thermalize a solid is of the order of tens of picoseconds, in 100 ns the sample achieves local thermodynamic equilibrium with a well-defined temperature that can be measured optically. Our DAC, a modified version of the Silvera and Wijngaarden cell [26], was cryogenically loaded. The DAC was then warmed to room temperature and removed from the cryostat. Two advantages exist for pulsed laser heating of hydrogen to reduce hydrogen diffusion into the confining elements: first hydrogen mainly diffuses during the hot pulse and this time is insufficient for important diffusive changes. Second, the gasket and diamonds do not get very hot [25]. The latter can be justified by the following argument. Although energy is flowing from the absorber to the hydrogen to the diamonds, the thermal time constants of the components and sample are very different. The metallic absorber is substantially smaller than the hydrogen sample and has a short thermal time constant compared to that of the hydrogen. The surface of the absorber warms during the pulse; the energy then thermalizes in the absorber in several microseconds to a much lower temperature than the peak temperature. The excess thermal energy from the absorber flows into the hydrogen at an even lower temperature and is conducted away into the high thermal conductivity diamonds until the DAC reaches ambient temperature. We use repetitive chains of laser pulses to enhance the signal-to-noise for the temperature measurement. The pulse repetition rate was 20 kHz, so that the duty cycle or on-off time was about 1:500, easily sufficient for the system to cool to ambient before the next pulse. The average laser power dissipated was a few hundred milliwatts for temperatures greater than $10^3$ K. After a heating cycle no detectable temperature rise of the DAC could be found by a touch test.

During a heating pulse the temperature rapidly rises to a peak and then falls off in several laser pulse widths [23]. If there is sufficient power, the melting will first take place at the peak temperature. In order to determine the peak temperature we measure the thermal radiation as a function of wavelength, $\lambda$. If $F_\lambda$ is the blackbody irradiance, then we measure

$$\overline{F}_\lambda = \int_0^\infty F_\lambda [T(t)] dt \,,$$

where T(t) is the temporal dependence of the temperature. The time average $\overline{F}_\lambda$ is dominated by the highest temperatures and resembles a Planck function. This curve is fit to a Planck function characterized by a temperature lower than the peak temperature, which we call the Planckian average temperature. By comparing $\overline{F}_\lambda$ and $F_\lambda$, a correlation table [23] can be developed for the measured spectrum to yield the peak temperature of $T(t)$, given the experimentally determined Planckian average temperature.

A critical technical problem for determining accurate temperatures by optical pyrometry is the determination of the optical transfer function, or the correction to the deformation of the broad band Planck function by the optical system, including the detector response. This is generally done by replacing the heated (glowing) source with a calibrated blackbody source of known spectral irradiance, to determine the transfer function. The deficiency of this technique is that the (glowing) absorber may be a gray-body emitter with a wavelength dependent emissivity, much different than the BB source. We use a new method that does not suffer from such uncertainties to determine the transfer function. We replaced the BB source with a platinum foil heated with our pulsed laser to a known temperature (just below the melting point of Pt, which is straightforward to determine [23]). This procedure circumvents the above problem as the same Pt foil is used as the absorber-emitter in the DAC. The transfer function was also corrected for the absorption and reflectivity of the two surfaces of the type IA diamond in the light path, taking into consideration the index of refraction of the hydrogen at the first surface [27]. Hydrogen itself is not expected to have any important absorption in our region of study.

Before carrying out the present experiment we confirmed that a melt line measured by the pulsed heating method agreed with existing measurements in the literature [24]. In that study the temperatures were quite high, of order 2000 K and the irradiance was measured with a sensitive CCD detector. However, for temperatures below ~1000 K there are very few photons in the visible spectrum as the irradiance falls off exponentially for short wavelengths and the BB irradiance has its peak in the IR where a CCD does not respond. Moreover the signal (compared to CW laser heating) is reduced by the duty cycle, a factor ~500. We changed our measurement procedure to using five narrow-band IR filters [28] and measured the irradiance with a liquid nitrogen cooled InSb detector. This was sensitive enough to determine the temperature down to ~700 K, where the signal-to-noise became noise limited. Pressure was determined using ruby fluorescence of the sample at ambient temperature and the new ruby pressure scale of Chijioke et al [29].

The solid-liquid melting transition was determined in two ways. First we observed laser speckle on a CCD detector that imaged our absorber. When the hydrogen melted we could see motion in the speckle pattern. If the sample was illuminated by an auxiliary laser operating continuously in the visible then speckle motion would not be observable because the melting only occurs at pulse peaks, about 1/500 of the time. However our CCD responded to the 1.065 micron pulsed laser radiation, which only illuminates the CCD during the laser pulse so that the duty cycle for observation of melting was of order 1. The observation of the plateau and motion of the laser speckle coincided. The footprint of the laser beam on the absorber was fairly uniform so that the sample temperature would have a small radial thermal gradient. The second method was to plot the temperature vs. the average laser power. With increasing power the temperature rises; there is a plateau at melting as shown in Fig. 2, as the energy goes into the heat of melting. In Fig.3 we show a typical set of data to determine the temperature of melting; the lower temperature is below the plateau; the higher on the plateau. Statistical errors of the temperature from the BB fit around the melt temperature are very small, around $\pm 4$ K. The smaller than usual uncertainty arises because with our method of using the platinum foil for the absorber and for determination of the optical transfer function, the emissivity drops out for the fitting procedure and one carries out a one-parameter fit rather than two (emissivity and temperature). Moreover, in unpublished work, we have shown that the emissivity of platinum is essentially independent of temperature in temperature region of our study. The dominant uncertainty arises from the determination of the melt temperature from the plateau curves, approximately $\pm 15$ K and a possible systematic error of ~5 K from the transfer function determination.

The melting line of hydrogen is shown in Fig. 4 and has a peak. Our lowest pressure measurement agrees well with the measurements of Gregoryanz et al. Our data exhibits a rather sharp peak at $64.7 \pm 4$ GPa and $1055 \pm 20$ K; the melt temperature falls off for higher pressures. The pressure we specify here is the ambient pressure, i.e. the pressure before heating and there may be a thermal effect (see ahead) during the short pulse that enhances the pressure at melting. Such an enhancement or correction varies slowly with pressure and would not suppress a maximum in the melting curve. The main source of uncertainty in the peak (ambient) pressure is due to the interval of data points. Our results establish a peak in the melting line in the vicinity of this point. We were not able to fit our data with the unexpectedly sharp peak to a Kechin melting curve, as was done by earlier researchers (compare our data to the curve of Bonev et al in Fig. 4). The calculation of the melt line depends on the space group of the solid and Bonev et al assumed the hcp structure. The unexpected sharpness of the melting line peak may be due to an as yet undetected solid-solid phase line intersecting the melt line; this might be an extension of a new phase recently detected by Baer, Evans, and Yoo [30].

There are two more interesting points to discuss. First, a preliminary FEA analysis [25], including thermoelastic coefficients, has shown that the pressure of the hydrogen at the surface of the absorber during a short heat pulse may be higher than the ambient pressure by of order 10-20%, depending on the pressure. This is not due to a shock wave, but due to the large thermal expansion coefficient of hydrogen. Thus, the ambient pressure we quote for the peak may be a lower bound and a detailed analysis of this effect is being pursued. Second, our experiment was concluded, not because the diamonds broke, but because the ruby signal for measuring the pressure continuously weakened as the pressures and temperatures were increased. When the load on the sample was increased to yield a pressure around 100 GPa, the ruby fluorescence could not be excited. Upon termination of the experiment the fluorescence of the ruby chips was studied; the chips recovered their fluorescence intensity within several hours. We believe that the high pressure loss of the ruby intensity was due to hydrogen diffusion that quenched the ruby fluorescence. Pressure can also be determined from the shift of the Raman active vibron in hydrogen or the phonon of the diamond in the vicinity of the diamond culet; calibrations for diamond by several researchers are in the literature [31]. Our attempt to use this pressure gauge was unsuccessful because the absorber had almost completely filled the gasket hole at the highest pressures, and the signal-to-noise for back scattering off of the absorber was too low to detect the Raman phonon edge, with similar problems for the hydrogen vibron that could also be used as a pressure gauge.

Pulsed laser heating easily extends to temperatures substantially higher than 1000 K for hydrogen. In fluid hydrogen in reverberating shock wave experiments a high electrical conductivity state is observed. However, the pressure temperature conditions reached in these experiments were far from the melt line [32]. If high-pressure low-temperature liquid metallic hydrogen exists, exciting new states of matter will come into existence. Future experiments should be carried out to extend the melting line to higher pressures. Time resolved Raman scattering experiments of fluid hydrogen should be able to detect the predicted liquid-liquid transition from molecular to atomic hydrogen.

We thank Anthony Papathanassiou for important experimental contributions in the earlier stages of this research project. We also thank Jonathan Nguyen, Jake Dweck, Anatoly Dmentyev, and Jieping Fang for experimental aid, and Jacques Tempere and Bill Nellis for useful discussions. We are grateful to Harvard University for providing us with support to enable the repair of our Nd:vanadate laser when it failed. Partial support for this research was provided by a special grant from Seychelles Footwear and Alby and Kim Silvera.

**Figure Captions**

Figure 1.  A sketch of the diamond culet flat region of the DAC for pulsed laser heating.  The platinum absorber sits on ruby chips to separate it from the diamond anvil surface. The ruby is also used for pressure determination.  The laser is focused beyond the absorber to broaden the footprint, resulting in a more uniform temperature.

Figure. 2.  Plots of the peak temperature vs. average laser power showing the rise in temperature and the plateaus when the hydrogen melts, for two different pressures.

Figure 3.  The blackbody radiation curves below and above the melt line for P=75 GPa.  For comparison, we also show the BB curve for 700 K, our approximate limit for temperature measurement.  Shown are the blackbody curves for the Planckian average temperatures of 921, 906,  and 866 K, yielding peak temperatures of 944, 925, and 886 K, respectively.

Figure. 4.  The experimental melting line of hydrogen showing our results along with earlier results at lower pressures.  Bonev et al fit their theoretical results to a Kechin curve and we show this curve (dashed line). We also show the calculated liquid-liquid phase line for dissociation of hydrogen in the melt.

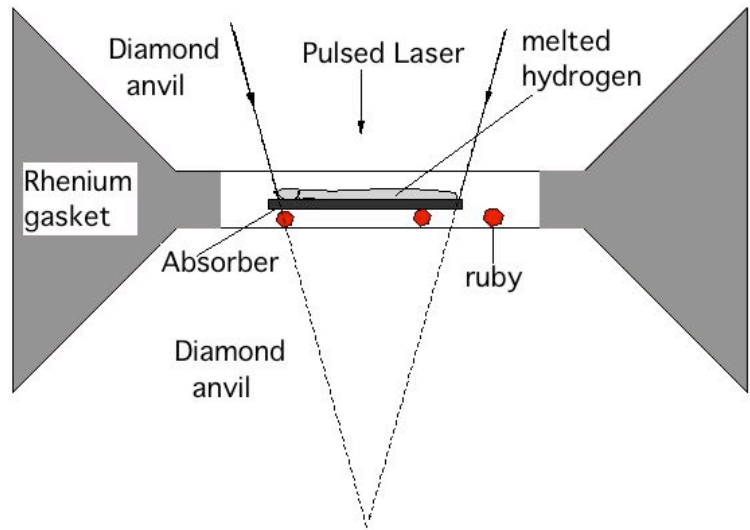

Fig. 1,

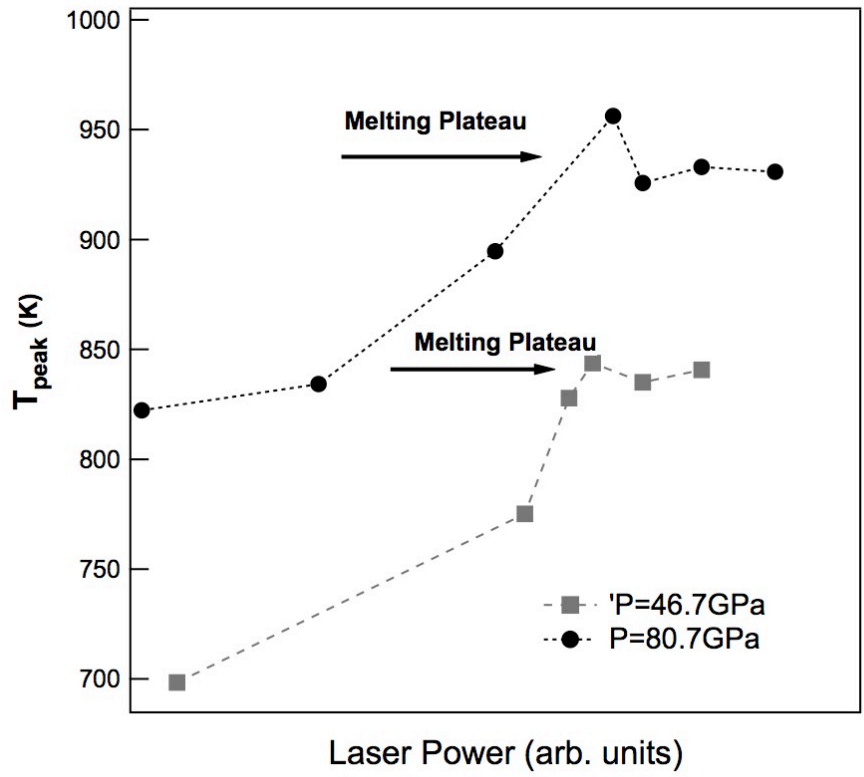

Fig.2

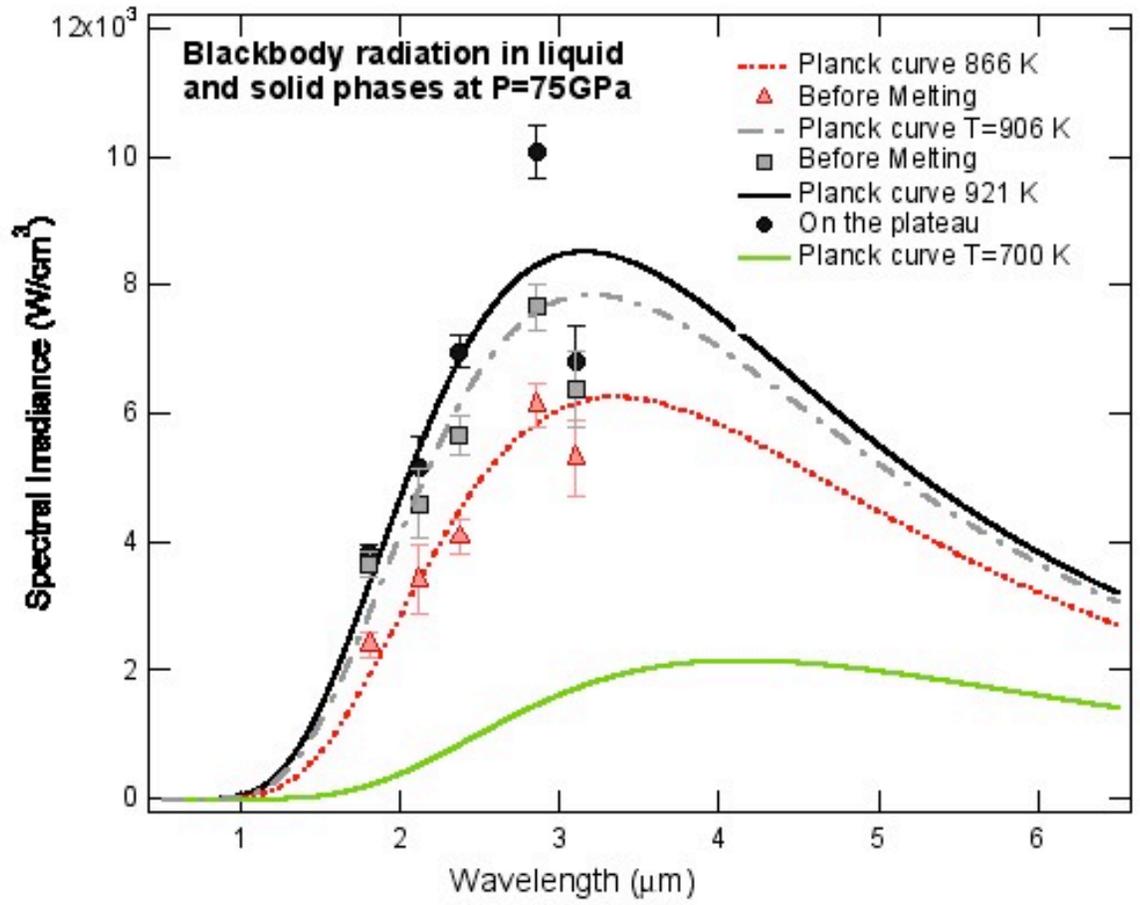

Fig. 3.

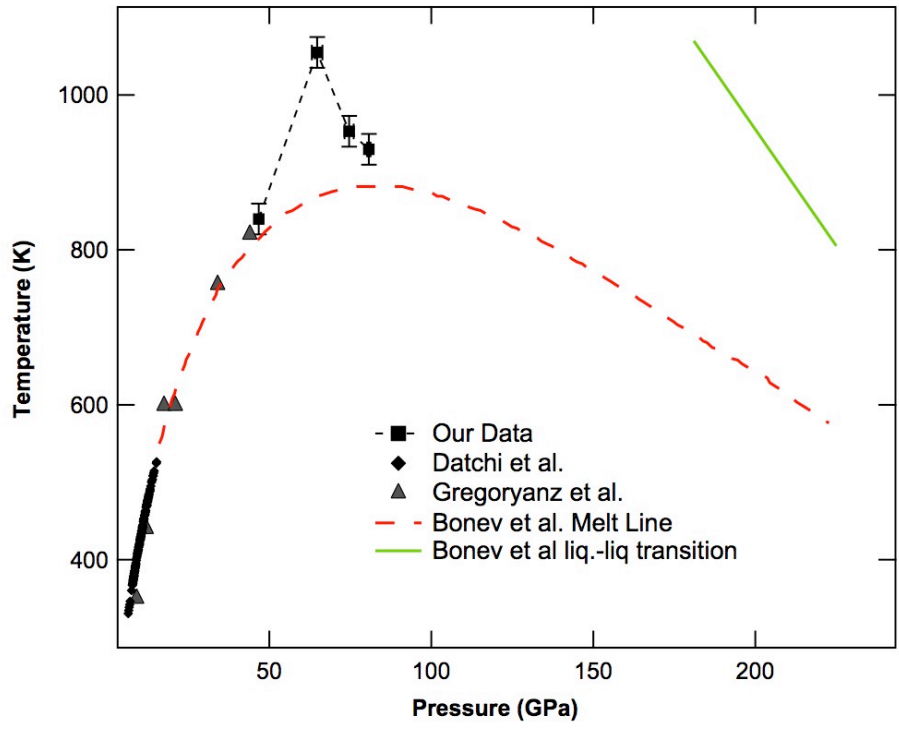

Fig. 4.